\pdfoutput=1

\documentclass[reprint,english,aps,prl,showpacs,superscriptaddress,preprintnumbers]{revtex4-1}
\usepackage{float}
\usepackage{amsmath}
\usepackage{amssymb}
\usepackage{braket}
\usepackage{graphicx}
\usepackage{xcolor}
\usepackage[colorlinks=true,urlcolor=blue,linkcolor=black,citecolor=black,breaklinks=true,bookmarks=false]{hyperref}

\hypersetup{pdfauthor={J. Oppermann, J. Straubel, K. Slowik, C. Rockstuhl},
pdftitle={Quantum Description of Radiative Losses in Optical Cavities}}

\newcommand{\ee}{\text{e}}
\newcommand{\ii}{\text{i}}
\renewcommand{\vec}[1]{\mathbf{#1}}

\begin{document}

\title{Quantum Description of Radiative Losses in Optical Cavities}
\author{J. Oppermann}
\email{jens.oppermann@kit.edu}
\affiliation{Institute of Theoretical Solid State Physics, Karlsruhe Institute of Technology, 76131 Karlsruhe, Germany}
\author{J. Straubel}
\affiliation{Institute of Theoretical Solid State Physics, Karlsruhe Institute of Technology, 76131 Karlsruhe, Germany}
\author{K. S{\l}owik}
\affiliation{Institute of Physics, Nicolaus Copernicus University, 87-100 Toru{\'n}, Poland}
\author{C. Rockstuhl}
\affiliation{Institute of Theoretical Solid State Physics, Karlsruhe Institute of Technology, 76131 Karlsruhe, Germany}
\affiliation{Institute of Nanotechnology, Karlsruhe Institute of Technology, 76131 Karlsruhe, Germany}

\begin{abstract}
We present, for the first time, the quantum mechanical description of light-matter interaction in the presence of optical cavities that are characterized by radiative losses. Unique to radiative losses is the unitary evolution and their full preservation of the coherence, in stark contrast to the usually considered dissipative losses. We elucidate the reduction of exact quantum electrodynamic equations to a form similar to the familiar Jaynes-Cummings model through the introduction and study of a new class of noise operators. The dynamics of this henceforth inherently dissipative model are then presented by formulating the resulting equations of motion. Furthermore, an input-output formalism is established, which provides a direct connection to the dynamics of output states accessible with detectors. The application-oriented cases of coherent and pulsed laser pumping are discussed as inputs. Finally the single-photon dynamics in an optical cavity with significant radiative loss - whose importance has to be contextualized in view of the prospects of light-matter interaction applications - are reviewed according to the proposed model. The formulation is kept as general as possible to emphasise the universal applicability to different implementations of quantum optical systems but from our own background we have an application in mind in the context of nanooptics.
\end{abstract}

\maketitle

\section{Introduction}
Quantum information science (QIS) employs quantum phenomena to elevate the processing and transfer of information to levels beyond the scope of classical physics~\cite{Cirac2017}. Consequently, in the past decades QIS has emerged as the stepping stone towards changing the very foundations of the information age~\cite{Schoelkopf2008, Kimble2008} and it constitutes nowadays a vibrant field of research~\cite{Zoller2005}. Similarly, quantum optics as one of QIS's mainstays shares these promising prospects.

The interaction of light and matter, that is encountered whenever an electromagnetic field impinges on some sort of media, is one of the most fundamental aspects of optics in general. As soon as discrete matter is considered, such as atoms or molecules instead of a bulk medium, it has been revealed to enable signal processing at the single-photon level~\cite{Imamog1999,Hennessy2007}. Cavity quantum electrodynamics (QED) models prototypical systems to study the quantum physical essentials of light-matter interaction~\cite{Cohen1989, Allen1987}. All these systems can be reduced to the phenotype of electromagnetic field modes usually sustained by some sort of cavity coupled to atomic transitions~\cite{Mandel1995, Scully1997}. The referential case of a single-mode quantized field coupled to a single transition defines the Jaynes-Cummings (JC) model~\cite{Jaynes1963, Cummings1965, Meystre2013, Berman1994}. Its widespread consideration towards problems of both fundamental and applied science is a clear indication for its success.

Although the notion of fundamental aspects such as nonclassical photon statistics~\cite{Planck1900} and nonclassical photon properties~\cite{Schrodinger1935} originated theoretically, experimental advancement has been the driving force that put the implementation of QED based applications in tangible proximity: from influencing the emission behavior of quantum systems via microcavities~\cite{Purcell1946}, to controllable sources of single photons~\cite{Kimble1977}, and generation of entangled photons~\cite{Shih1988,Ou1988}. More recent developments were generally focussed on controlling transition properties in quantum systems through tailored electromagnetic field modes~\cite{Miller2005,Novotny2011,De2012,Gambino2014}. In this context, optical cavities~\cite{Yamamoto1995,Vahala2003,Schuller2010} have proven to be a versatile and powerful tool to exercise control over the properties of the field modes~\cite{Hood2001,Kelly2003,Garcia2009}. Consequently, a variety of investigations targeted e.g. the generation of squeezed states of light~\cite{Slusher1985,Wu1986}, single photons~\cite{Kuhn2002,Schietinger2009,Filter2014,Lee2011,Esteban2010,Straubel2016}, entangled photons~\cite{Plenio2002,Dousse2010,Oka2013,Hou2014,Oka2015,Straubel2017}, and nonclassical light in general~\cite{Maksymov2010,Slowik2014} - all relying on features of the optical cavity.

\begin{figure}[t!]
\begin{center}
\includegraphics[width=.45\textwidth]{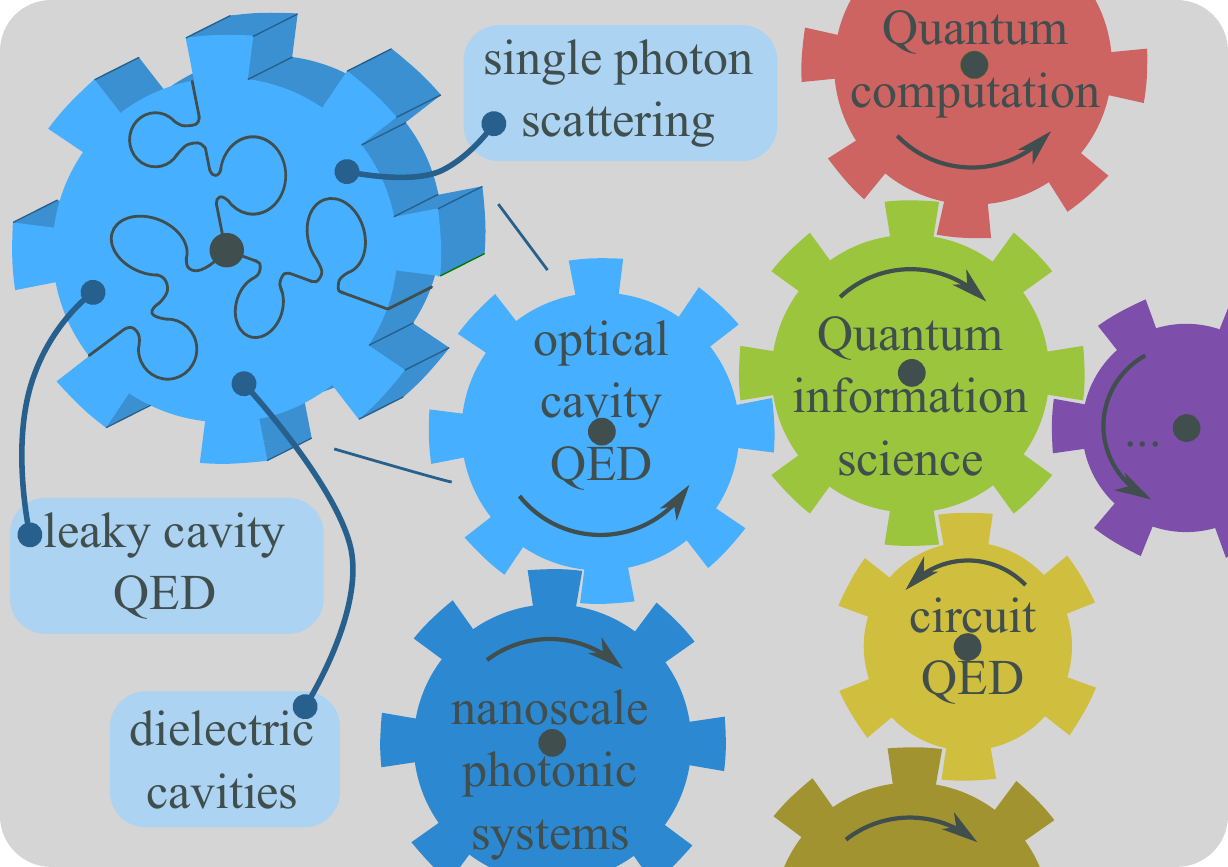}
\end{center}
\caption{Subsumption of the treatment introduced in this work into the research activities and goals. Our work eventually contributes to the gear shown in the upper left corner, but it has an impact on a multitude of further research areas.}
\label{fig:schematic}
\end{figure}

However, these results do not stand on their own as autotelic applications, but are rather motivated by the greater goal of expanding the cavity QED tool box and to ultimately culminate to the implementation of quantum computation via QIS [see Fig.~\ref{fig:schematic}]. This purpose immediately leads to the aspect of cavity mediated coupling regimes~\cite{Andreani1999,Hummer2013,Slowik2013,Esteban2014,Chikkaraddy2016} in light-matter interaction and quantum coherence~\cite{Vedral1998,Mabuchi2002,Sillanpaa2007}. These facets have been widely discussed as well, including their quantum information prospects~\cite{Cirac1997,Bose1999,Pachos2002,Ginzburg2010,Jacob2011,Tame2013}.

At this point it is crucial to highlight that any light-matter interaction scenario utilizing optical cavities hinges on the radiation emitted by the cavity. The emitted photons are the carriers of the desired features or states and shall be used to ultimately encode information. But so far, the modeling of the emission process itself has been assumed to be equivalent to the coupling to a thermal bath~\cite{Ginzburg2016}. This is not true. As outlined further below in slightly more detail, in contrast to dissipation to a thermal bath, radiative losses don't lead to the thermalization of energy and the corresponding loss of information. Instead the electromagnetic energy is transferred into the far field, where full information can be retrieved with photodetectors. Besides the inability to actually describe the measurement process, this urgently prompts for an adapted description of the light-matter interaction in the presence of a cavity with radiation losses.

\begin{figure}[h!]
\begin{center}
\includegraphics[width=.45\textwidth]{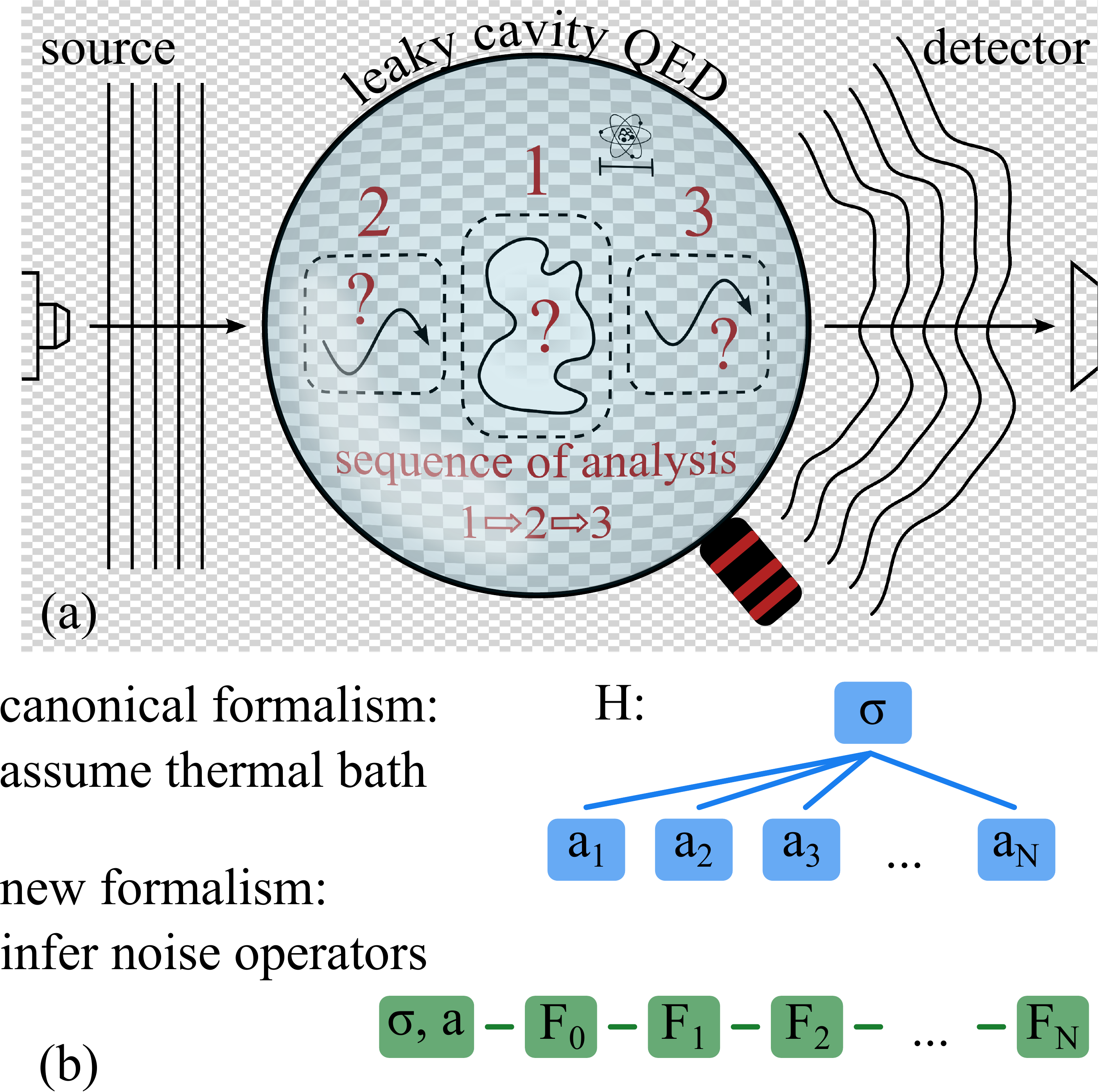}
\end{center}
\caption{(a) System under investigation: An arbitrary optical cavity is exposed to arbitrary incident illumination, which results in emission into the far field. (b) Methodological change in formulation of emission: Coupling to thermal bath is replaced by new coupling to chain of noise operators.}
\label{fig:systematic}
\end{figure}

To close this fundamental methodological gap we propose a formulation via a set of operators introduced in the following and derived from first principles. Incipiently we would like to stress that we do not propose to replace one assumption with another, but rather based on an analysis of the internal dynamics, we inferred a new structure of the emission formulation from the methodology we introduce here [see Fig.~\ref{fig:systematic}(b)]. With optical cavities universally relying on radiative losses to function as potential QIS devices, we propose a conscientious formalistic incorporation of the radiative loss aspect of light-matter interaction modelling in cavities into the well-established theory. This entails the integration of the emission of photons into the measurable far field into the rigorous description of the unitary internal dynamics of the cavity [see Fig.~\ref{fig:systematic}(a) for the general setup we are looking at].

In lossless cavity QED, the temporal evolution of the joint system consisting of light and matter is unitary and can therefore be described by a Hamiltonian. The simplest of such systems consists of a single electromagnetic mode and an atom, where the interaction between both is described in dipole approximation. Such a system is described by the Rabi Hamiltonian, which in many cases can be further simplified to yield the Jaynes-Cummings Hamiltonian.

However, since there exist no lossless systems in reality, the description of realistic systems requires that losses in the system are somehow taken into account. This is usually achieved by including the coupling of system operators to a bath of other excitations, e.g. phonons. Since realistic systems contain a large number of such unwanted excitations, the energy of the system is eventually distributed somewhat evenly across all of them and therefore effectively lost. If one thinks about optical processes, then it is clear that absorption losses can be modeled in this way~\cite{Ginzburg2016rev}.

In this work, however, we consider the case of radiative losses. In contrast to absorption losses, radiative losses do not lead to the loss of electromagnetic energy. Instead, the energy is just displaced over time to a place far away from the system of interest. Applying the procedure outlined above would therefore require to place a perfect absorber around the cavity, effectively leading to the far field being integrated out. Since most experimental setups place their detectors into the far field, this would mean that we lose information about the observables of our theory. For this reason we introduce here an alternative formalism that splits the electromagnetic field itself up into one system and multiple bath operators. This leads to dissipative dynamics and provides insights concerning the relation between scattering and cavity modes.

This work is organized as follows. In Sec.~\ref{sec:eom} the electromagnetic cavity and bath operators are defined and their equations of motion derived. The bath operators are then integrated out to yield closed equations containing only the operators of interest. Section~\ref{sec:io} establishes the connection between cavity and far field dynamics, allowing for the formulation of an input-output scheme. In Sec.~\ref{sec:dynamics} the formalism is used to retrieve a generalized form of the experimentally well-established Jaynes-Cummings model, which can also describe the far field dynamics. We summarize our findings in Sec. \ref{sec:conclusion}. The Appendices contain some of the more involved calculations needed throughout this work.

\section{The Cavity}
\label{sec:eom}
We consider the problem of a quantized electromagnetic field coupled to a single two-level system (TLS) in electric dipole approximation. We further assume the presence of a localized and lossless dielectric structure giving rise to classical light scattering. The Hamiltonian of the system can be found in many textbooks on quantum optics and reads \cite{Vogel1994}
\begin{align} \label{eq:exact_hamiltonian}
H =& \sum_{\lambda}\int d^3k \hbar\omega_{\vec{k}} a_{\vec{k},\lambda}^\dagger a_{\vec{k},\lambda} + \hbar\omega_{a}\frac{\sigma_z}{2} \nonumber\\
& + \sum_{\lambda}\int d^3k\left(\vec{E}_{\vec{k},\lambda}(\vec{r}_a)\cdot \vec{d} a_{\vec{k},\lambda}^\dagger \sigma_- + h.c.\right),
\end{align}
where $\omega_{\vec{k}} = c_0 \left|\vec{k}\right|$, $c_0$ is the speed of light in vacuum, $\vec{d}$ and $r_a$ are the transition dipole moment and spatial position of the TLS, $\sigma_z$ is a Pauli matrix, $\sigma_-$ is the corresponding Pauli lowering operator and $a_{\vec{k}, \lambda}$ are photonic annihilation operators. For the rest of this work we drop the mode index $\vec{k}$ from $\omega_{\vec{k}}$, making the dispersion relation implicit. The electromagnetic field modes are of the form \cite{Oppermann2017}
\begin{equation} \label{eq:eigenmodes}
\vec{E}_{\vec{k},\lambda}(\vec{r}) = \sqrt{\frac{\hbar\omega}{(2\pi)^32\epsilon_0\epsilon_b}} \hat{\vec{e}}_{\vec{k},\lambda} \ee^{\ii \vec{k} \vec{r}} + \vec{E}_{\vec{k},\lambda}^{(s)}(\vec{r}),
\end{equation}
where $\epsilon_b$ is the permittivity of the background medium and $\vec{E}_{\vec{k},\lambda}^{(s)}(\vec{r})$ is the scattered field, which arises due to the spatially inhomogeneous dielectric function $\epsilon(\vec{r}, \omega)$. The photonic operators satisfy the usual harmonic oscillator commutation relations
\begin{equation}
\left[a_{\vec{k},\lambda}, a^\dagger_{\vec{k'}, \lambda'} \right] = \delta(\vec{k} - \vec{k'})\delta_{\lambda \lambda'},
\end{equation}
with all other commutators vanishing. The description offered by the Hamiltonian in Eq.~\eqref{eq:exact_hamiltonian} is exact and general, but usually not tractable. Luckily, many problems of practical interest feature electromagnetic resonances, i.e. the electromagnetic field at the position of the TLS only takes on appreciable values over a set of comparatively narrow frequency ranges. In the following we will assume the existence of a single resonant electromagnetic mode in the vicinity of the transition frequency $\omega_a$ of the TLS. All other modes will be assumed to be too far detuned to yield any significant contribution and therefore can be ignored. By further assuming a Lorentzian frequency dependence, we write
\begin{equation} \label{eq:lorentzian_mode}
\vec{E}_{\vec{k},\lambda}(\vec{r}_{a}) = \vec{E}_{0} \delta_{\lambda, \lambda_0} \sqrt{\frac{\Gamma}{2\pi}}\frac{g(\hat{\vec{k}})}{\omega-\omega_0-\ii\Gamma/2},
\end{equation}
where $\omega_0$ is the central frequency and $\Gamma$ is the linewidth of the mode and $g(\hat{\vec{k}})$ describes the angular dependence. The Kronecker delta $\delta_{\lambda, \lambda_0}$ signifies that there is indeed only one electromagnetic mode and not two degenerate ones of different polarization. This constitutes no limitation since in practice one can describe the polarizations in the coupled-uncoupled basis. Since only one polarization couples to the TLS we shall drop polarization indices $\lambda$ from here on. In the following, we will show that the above assumptions allow us to reduce the exact Hamiltonian to an approximate but tractable form.										
Taking Eq.~\eqref{eq:lorentzian_mode} into account, we can derive evolution equations in the Heisenberg picture for the operators from Eq.~\eqref{eq:exact_hamiltonian} that read as
\begin{align} \label{eq:exact_heisenberg_a}
\dot{a}_{\vec{k}} =& -\ii\omega a_{\vec{k}} \nonumber\\
& -\ii\frac{\vec{E}_{0}\cdot\vec{d}}{\hbar}\sqrt{\frac{\Gamma}{2\pi}}\frac{g(\hat{\vec{k}})}{\omega-\omega_0-\ii\Gamma/2}\sigma_- ,\\
\label{eq:exact_heisenberg_s}
\dot{\sigma}_- =& -\ii\omega_{a}\sigma_- \nonumber\\
& +\ii\frac{\vec{E}^*_{0}\cdot\vec{d}^*}{\hbar}\int d^3k \sqrt{\frac{\Gamma}{2\pi}}\frac{g^*(\hat{\vec{k}})}{\omega-\omega_0+\ii\Gamma/2}\sigma_za_{\vec{k}}.
\end{align}
Inspection of Eq.~\eqref{eq:exact_heisenberg_s} motivates the following definition of a resonant mode annihilation operator:
\begin{eqnarray} \label{eq:cavity_mode_definition}
a &:=& \int \frac{c_0^{3/2}d^3k}{\sqrt{G}\omega}\sqrt{\frac{\Gamma}{2\pi}}\frac{g^*(\hat{\vec{k}})}{\omega-\omega_0+\ii\Gamma/2}a_{\vec{k}},\\
\label{eq:cavity_mode_normalization}
G &:=& \int d\Omega_k \left|g(\hat{\vec{k}})\right|^2,
\end{eqnarray}
where $d\Omega_k = \sin(\theta_k)d\theta_kd\varphi_k$ denotes integration over solid angles. Please note that the normalization constant $G$ in Eq.~\eqref{eq:cavity_mode_normalization} is chosen in such a way that the harmonic oscillator commutation relations are satisfied:
\begin{equation}
\left[a, a^\dagger\right] = 1,
\end{equation}
and all other commutators vanishing. The original Heisenberg Eqs.~\eqref{eq:exact_heisenberg_a} and \eqref{eq:exact_heisenberg_s} can now be rewritten in terms of the newly defined resonant mode operators. The process is detailed in Appendix \ref{app:as_deriv} and the results read
\begin{align}
\label{eq:effective_heisenberg_a}
\dot{a} &= (-i\omega_0-\Gamma/2)a - \ii\kappa\sigma_- -\ii F_0, \\
\label{eq:effective_heisenberg_s}
\dot{\sigma}_- &= -\ii\omega_{a}\sigma_-+\ii\kappa^*\left(1-\ii\frac{\Gamma}{2\omega_0}\right) \sigma_za_{} +\ii\frac{\kappa^*}{\omega_0} \sigma_z F_0,
\end{align}
where the effective light-matter coupling constant $\kappa$ is defined as
\begin{align} \label{eq:coupling_constant}
\kappa &= \frac{\sqrt{G}}{c_0^{3/2}}\omega_0\frac{\vec{E}_{0}\cdot\vec{d}}{\hbar} \nonumber \\
&= \sqrt{\frac{G}{4\pi}} \frac{\pi}{c_0^{3/2}}\omega_0\sqrt{2\Gamma}\frac{\vec{E}_{\text{max}}\cdot\vec{d}}{\hbar},
\end{align}
with $\vec{E}_{\text{max}}$ being the field strength at resonance. To the best of our knowledge this is the first time that the coupling strength between an open cavity mode and a quantum emitter has been calculated from first principles rather than from phenomenological considerations. The new operator $F_0$ appearing in Eqs.~\eqref{eq:effective_heisenberg_a} and \eqref{eq:effective_heisenberg_s} belongs to a family of operators defined as
\begin{equation} \label{eq:fluctuation_definition}
F_n := \int \frac{d^3k}{\omega}\frac{c_0^{3/2}}{\sqrt{G}}\sqrt{\frac{\Gamma}{2\pi}}g^*(\hat{\vec{k}}) \left(\omega - \omega_0\right)^n a_{\vec{k}}.
\end{equation}
Comparing Eqs.~\eqref{eq:cavity_mode_definition} and \eqref{eq:fluctuation_definition}, we notice that the operators $F_n$ are not associated with the resonance mode at $\omega_0$, but rather with a broad range of frequencies. For this reason we will call $F_n$ noise operators from here on. Please note that these noise operators were retrieved without having introduced a thermal bath.

In order to construct a closed system of equations, we need equations of motion for $F_n$. The derivation of these is detailed in Appendix \ref{app:fn_deriv}. The results read
\begin{equation} \label{eq:effective_heisenberg_f}
\dot{F}_n = -\ii \omega_0 F_n - \ii F_{n+1}.
\end{equation}
The set of equations \eqref{eq:effective_heisenberg_a}, \eqref{eq:effective_heisenberg_s}, and \eqref{eq:effective_heisenberg_f} is closed and can therefore be used in their current form to describe the system dynamics. In the spirit of the theory of open quantum systems, however, we wish to find a set of equations that only contains the system operators $a$ and $\sigma_-$ as dynamic quantities.

As is detailed in Appendix \ref{app:fn_sol}, the equations of motion of the noise operators \eqref{eq:effective_heisenberg_f} can be formally solved without further approximations to yield
\begin{align} \label{eq:solution_f}
F_n(t) &= \ee^{-\ii\omega_0 t} \sum_{m = 0}^\infty \frac{(-\ii t)^m}{m!}F_{m+n}(0),
\end{align}
where the form of $F_{m}(0)$ defines the type of illumination, as can be seen from Eq.~\eqref{eq:fluctuation_definition} and the examples of Section \ref{sec:io}.
Please note that the operators $F_n(t)$ can be interpreted as input parameters, as Eq.~\eqref{eq:solution_f} tells us that there exists no backaction from the system. Insertion of Eq.~\eqref{eq:solution_f} into Eqs.~\eqref{eq:effective_heisenberg_a} and \eqref{eq:effective_heisenberg_s} now yields a closed set of equations of motion for the operators $a$ and $\sigma_-$, where the lowest order input operator serves as a pump term
\begin{align}
\label{eq:complete_heisenberg_a}
\dot{a}(t) =& (-i\omega_0-\Gamma/2)a(t) - \ii\kappa\sigma_-(t) - \ii F_0(t), \\
\label{eq:complete_heisenberg_s}
\dot{\sigma}_-(t) =& -\ii\omega_{a}\sigma_-(t) +\ii\kappa^*\left(1-\ii\frac{\Gamma}{2\omega_0}\right) \sigma_z(t)a(t) \nonumber \\
& + \ii\frac{\kappa^*}{\omega_0} \sigma_z(t) F_0(t). 
\end{align}
Using Eqs.~\eqref{eq:complete_heisenberg_a} and \eqref{eq:complete_heisenberg_s} allows one to find equations of motion for all observables of the system, i.e. expectation values of arbitrary operators. The inital values of operators containing $a$ and $F_n$ can be inferred from the initial state of the quantized electrodynamic field using the definitions \eqref{eq:cavity_mode_definition} and \eqref{eq:fluctuation_definition}. The structure of these equations resembles the one presented conceptually in Fig.~\ref{fig:systematic}(b). Instead of the simultaneous coupling of the system to a larger number of bath operators, the operators that describe the evolution of our actual system is only coupled to one noise operator. In a sequential type of process each noise operator then couples to the next.

We have now succeeded not only at describing the internal quantum dynamics of a cavity with radiative losses, but also at linking them to the external field via the noise operator $F_0(t)$. In the following section this link will be used to formulate an input-output scheme capable of describing real experimental setups.

\section{Input-Output Formalism}
\label{sec:io}

\begin{figure}
\includegraphics[width=.25\textwidth]{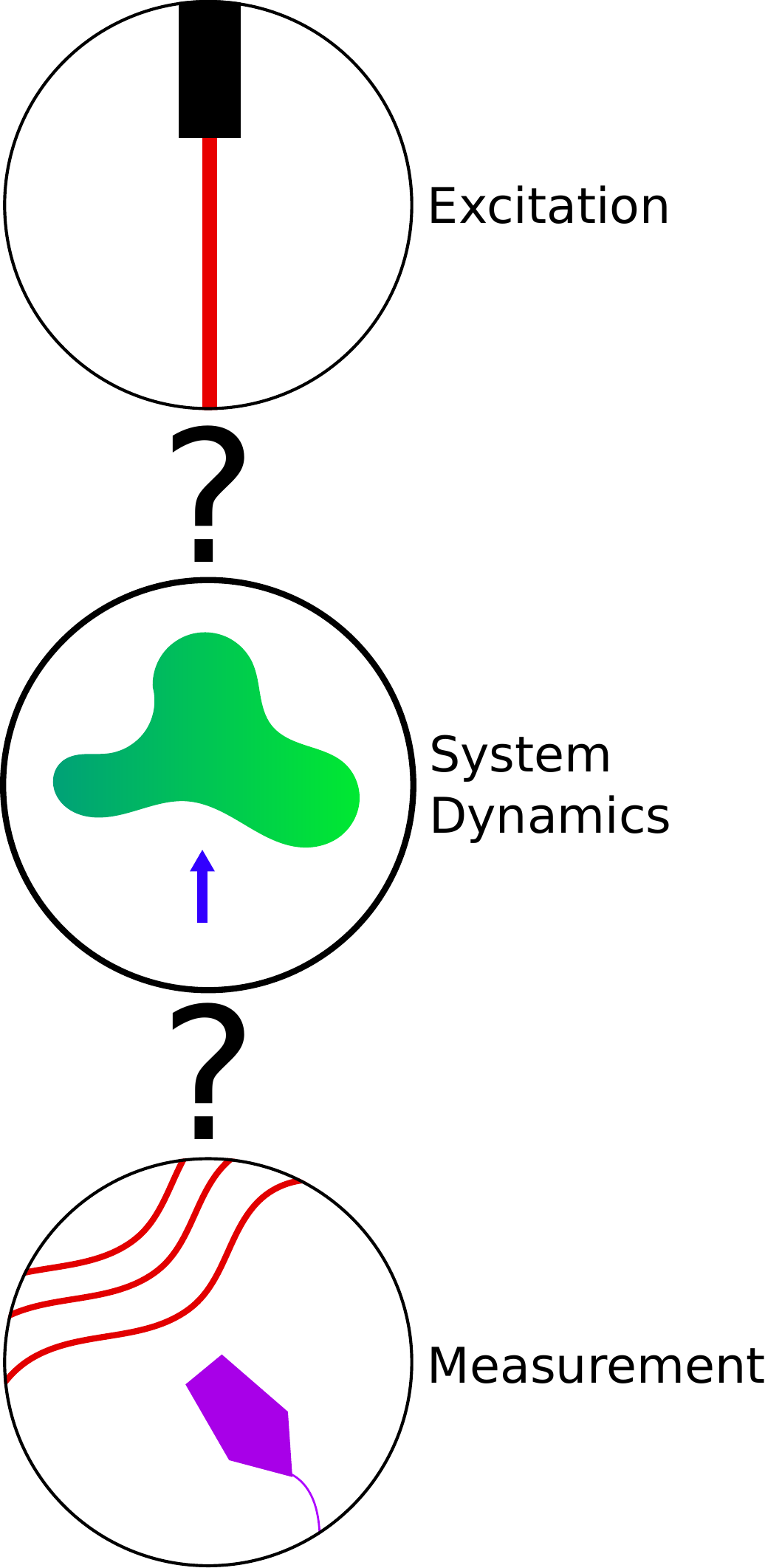}
\caption{Schematic depiction of an experimental setup. The system is excited by a beam of light, the interaction between cavity mode and matter takes place and finally the outgoing radiation is measured in the far field.}
\label{fig:io_scheme}
\end{figure}

At this point we succeeded in describing the internal dynamics of a cavity-matter system in a tractable and rigorous manner. However, this is not yet sufficient to make predictions about actual experiments. A typical quantum optical experiment consists of probing an optical system with a beam of light and measuring the outgoing radiation. This process is schematically shown in Fig.~\ref{fig:io_scheme}. In the following, we will establish a quantitative relation between the different parts of the systems.

\subsection{Input Operators}
\label{sec:input}

Most experimental illumination schemes in optics use a light beam with a waist diameter much larger than the system under consideration. For all practical purposes such a light beam can be considered as a plane wave with the wave vector pointing along the beam axis. For this reason, the following subsections describe how an illumination with coherent plane waves can be described by means of the noise operator $F_0(t)$.

\subsubsection{Continuous pumping}

One of the most common pumping schemes is excitation by a continuous laser beam, i.e. the incident light is monochromatic, coherent, and polarized. In terms of the scattering eigenmode operators this means
\begin{align}
\braket{a_{\vec{k}}(t=0)} &= \alpha_P \delta(\vec{k}-\vec{k}_P),
\end{align}
where $\alpha_P$ is the amplitude and $\vec{k}_P$ is the wave vector of the laser beam. Using Eq.~\eqref{eq:fluctuation_definition} one finds
\begin{align} \label{eq:coherent_fn}
\braket{F_n(0)} &= C (\omega_P-\omega_0)^n, \\
C &:= \frac{c_0^{3/2}}{\omega_P}\sqrt{\frac{\Gamma}{2\pi G}}g(\vec{\hat{k}}_P) \alpha_P. \nonumber
\end{align}
Substitution of Eq.~\eqref{eq:coherent_fn} into the expectation value of Eq.~\eqref{eq:solution_f} now yields
\begin{align} \label{eq:coherent_pump}
\braket{F_0(t)} &= C \ee^{-\ii\omega_0 t} \sum_{n=0}^\infty \frac{(-i(\omega_P-\omega_0)t)^n}{n!} \nonumber \\
&= C \ee^{-\ii\omega_0 t} \ee^{-\ii(\omega_P - \omega_0)t} \nonumber \\
&= C \ee^{-\ii\omega_P t}.
\end{align}
Equation \eqref{eq:coherent_pump} implies that the equation of motion for $a$ contains a pump term of constant amplitude, which oscillates at the pump laser frequency. Terms of this form have been routinely employed when discussing driven quantum systems, but now we actually have the means to quantify the relation between laser intensity and pump strength.

\subsubsection{Pulsed pumping}

Assume now that the laser used to pump the system is not continuous, but pulsed. This means that a range of frequencies centered around the laser frequency is excited according to
\begin{align}
\braket{a_{\vec{k}}} = \alpha_P \delta(\vec{\hat{k}} - \vec{\hat{k}}_P)\ee^{-\Delta^2(\omega-\omega_0)^2}.
\end{align}
Inserting this into the expectation value of Eq.~\eqref{eq:fluctuation_definition} leads to
\begin{align} 
\braket{F_n(0)} &= C \int_0^\infty d\omega \omega (\omega-\omega_0)^n \ee^{-\Delta^2(\omega-\omega_0)^2}, \\
C &= c_0^{3/2}\sqrt{\frac{\Gamma}{2\pi G}}g^*(\vec{\hat{k}}_P) \alpha_P. \nonumber
\end{align}
The frequency integral can be evaluated after extending the lower integration boundary to $-\infty$ to yield \cite{Owen1980}
\begin{align} \label{eq:pumped_fn}
\braket{F_{2n}(0)} &= C\omega_0\sqrt{\pi} \frac{(2n-1)!!}{2^n\Delta^{2n+1}} \\
\label{eq:pumped_fn_zero}
\braket{F_{2n+1}(0)} &= C\sqrt{pi} \frac{(2n+1)!!}{2^{n+1}\Delta^{2n+3}},
\end{align}
where the double factorial is defined as
\begin{align}
n!! &= \prod_{k=0}^{\lceil{n/2}\rceil-1}(n-2k).
\end{align}
Insertion of Eqs.~\eqref{eq:pumped_fn} and \eqref{eq:pumped_fn_zero} into the expectation value of Eq.~\eqref{eq:solution_f} now leads to
\begin{widetext}
\begin{align}
\braket{F_0(t)} &= \ee^{-\ii\omega_0 t} \sum_{n=0}^{\infty} \frac{(-\ii t)^n}{(2n)!} \braket{F_{2n}(0)} + \ee^{-\ii\omega_0 t} \sum_{n=0}^\infty \frac{(-\ii t)^{2n+1}}{(2n+1)!} \braket{F_{2n+1}(0)} \nonumber \\
&= \ee^{-\ii\omega_0 t} \frac{\sqrt{\pi} C \omega_0}{\Delta} \sum_{n=0}^\infty \left(-\frac{t^2}{2\Delta^2}\right) \frac{(2n-1)!!}{(2n)!} + \ee^{-\ii\omega_0 t} \sqrt{\pi}C \left(\frac{-\ii t}{2\Delta^3}\right) \sum_{n=0}^\infty \left(-\frac{t^2}{2\Delta^2}\right) \frac{(2n+1)!!}{(2n+1)!} \nonumber \\
&= \ee^{-\ii\omega_0 t} \frac{\sqrt{\pi} C \omega_0}{\Delta} \sum_{n=0}^\infty \frac{1}{n!} \left(-\frac{t^2}{(2\Delta)^2}\right) + \ee^{-\ii\omega_0 t} \sqrt{\pi}C \left(\frac{-\ii t}{2\Delta^3}\right) \sum_{n=0}^\infty \frac{1}{n!} \left(-\frac{t^2}{\left(2\Delta\right)^2}\right) \nonumber \\
&= \ee^{-\ii\omega_0 t} \frac{\sqrt{\pi} C \omega_0}{\Delta} \left[1-\ii\frac{t}{2\omega_0\Delta^2}\right] \ee^{\left(-\frac{t^2}{(2\Delta)^2}\right)} \approx \frac{\sqrt{\pi} C \omega_0}{\Delta} \ee^{\left(-\frac{t^2}{(2\Delta)^2}\right)} \ee^{-\ii\left(\omega_0 - \frac{1}{2\omega_0\Delta^2}\right)t},
\end{align}
\end{widetext}
where in the last step $\omega\Delta << 1$ was assumed. We therefore once again arrive at the expected result, namely that the pump term gains a Gaussian envelope with a temporal spread equal to the inverse of the frequency spread.

\begin{figure}
\includegraphics[width=.25\textwidth]{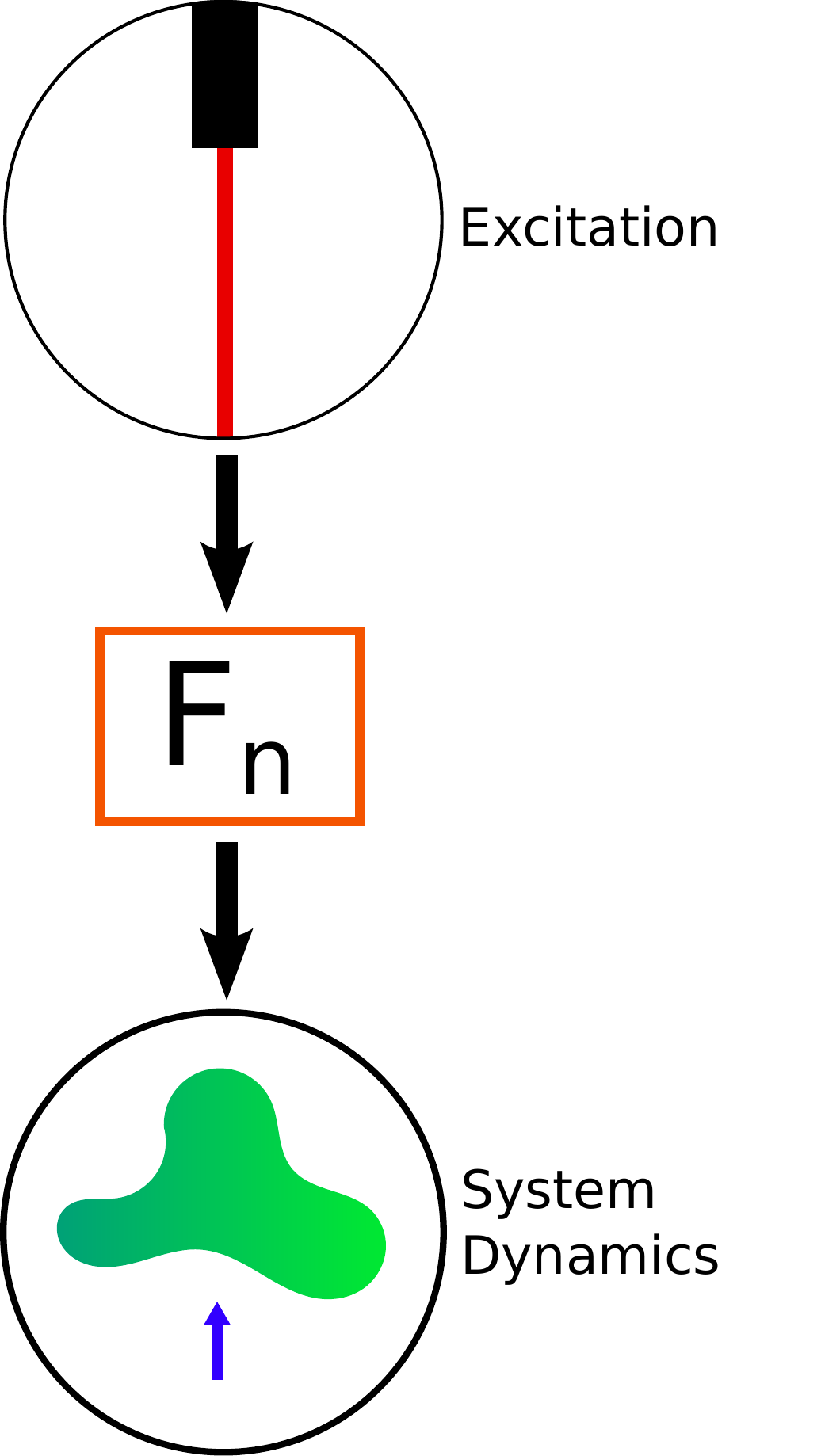}
\caption{The noise operators are defined in terms of scattering modes and can therefore bridge the gap between far field excitation and internal dynamics.}
\label{fig:input}
\end{figure}

The above examples demonstrate, that the noise operators are fully capable of describing arbitrary input schemes in a quantitative manner. They form a link between the excitation and internal dynamics of the optical system, as is illustrated in Fig.~\ref{fig:input}.

\subsection{Output Operators}
\label{sec:output}

Up to this point we have only been concerned with the internal dynamics of the system under external irradiation. But in order to describe actual experiments, we also need to consider the dynamics of output states, that are experimentally accessible via detectors. If all detectors are placed in the far field,  have sufficiently small apertures, and are sensitive to only a narrow frequency range, the output modes we are interested in are of the form
\begin{align} \label{eq:output_modes}
\vec{E}^{\text{(out)}}_{\vec{k},\lambda}(\vec{r}) = C_{\vec{k},\lambda} \hat{\vec{e}}_{\vec{k},\lambda} \ee^{\ii \vec{k} \vec{r}} + \vec{E}_{\vec{k},\lambda}^{(i)}(\vec{r}),
\end{align}
where $\vec{E}_{\vec{k},\lambda}^{(i)}(\vec{r})$ only contains incoming field components. A solution to Maxwell's equations of the form \label{eq:output_mode} describes a complicated scenario, in which the scattering responses of all incident fields interfere destructively in all but one spacial direction. This leads to a single plane wave as an outgoing field.

\begin{figure}
\includegraphics[width=.25\textwidth]{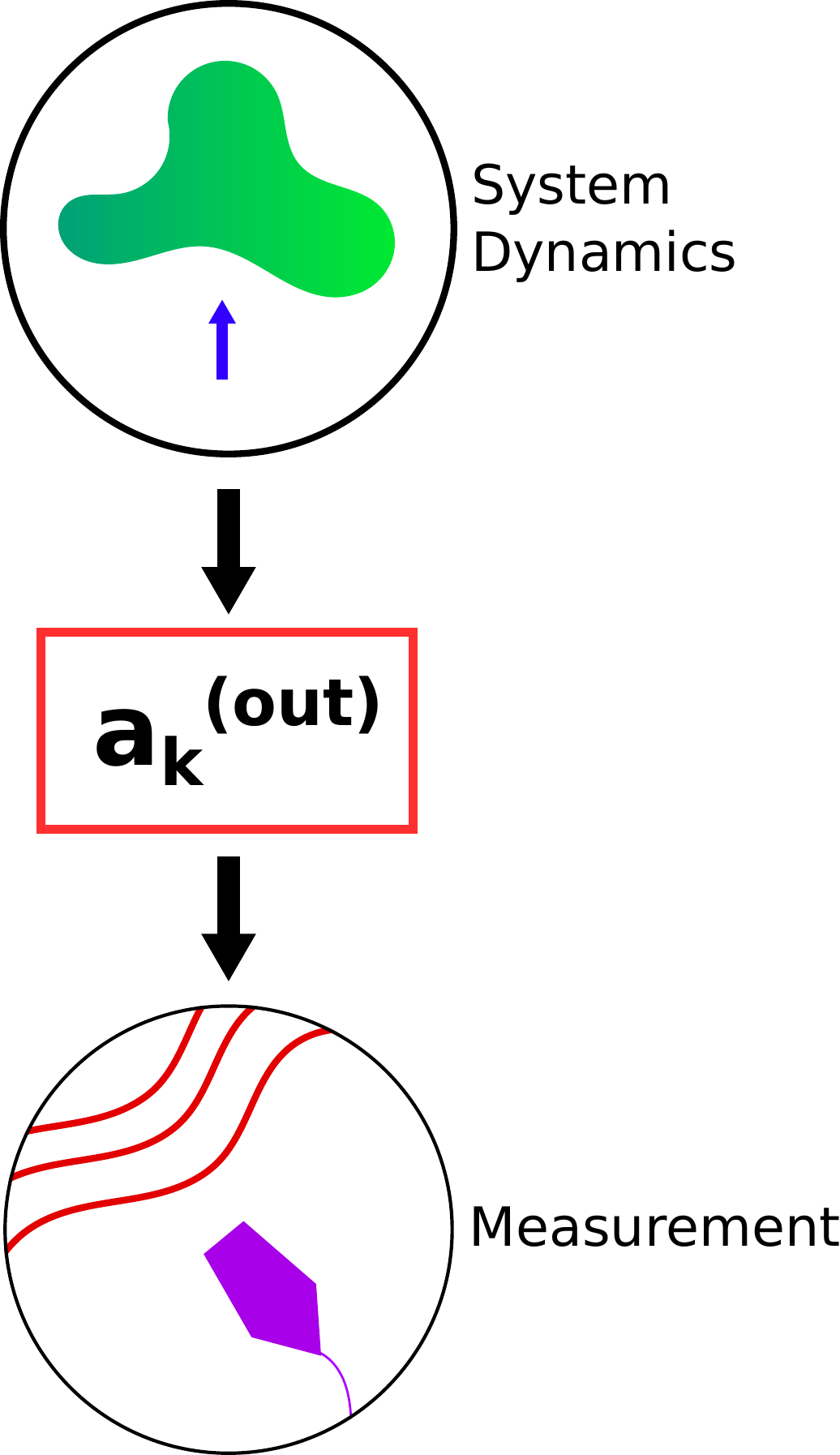}
\caption{The time reversed scattering modes have only a single outgoing Fourier component and therefore describe the field radiated away into a specific direction.}
\label{fig:output}
\end{figure}

One should of course ask, whether solutions of the form \eqref{eq:output_modes} exist and how one can find them. In order to answer these questions, we first recall that the macroscopic Maxwell equations are invariant under time-reversal, if no absorption losses are present \cite{Rachidi2013}. This means that the system Hamiltonian is invariant under the antiunitary time reversal operator $T$ \cite{Messiah1961}:
\begin{align}
THT &= H.
\end{align}
We can therefore exchange the electric field for its time-reversed counterpart, without changing the structure of the Hamiltonian. Therefore, the Heisenberg equations of motion also keep their form under time-reversal. The time-reversed electric field operator reads
\begin{align} \label{eq:time_reversed_electric_field}
T\vec{E}(\vec{r})T &= \int d^3k \left[T\vec{E}_{\vec{k},\lambda}(\vec{r}) a_{\vec{k}}T + h.c. \right] \nonumber \\
&= \int d^3k \left[\vec{E}^*_{\vec{k},\lambda}(\vec{r}) Ta_{\vec{k}}T + h.c. \right] \nonumber \\
&= \int d^3k \left[\vec{E}^*_{-\vec{k},\lambda}(\vec{r}) Ta_{-\vec{k}}T + h.c. \right].
\end{align}
We now define the time-reversed photon operators
\begin{align}
a^{\text{(out)}}_{\vec{k}} &= Ta_{-\vec{k}}T, \nonumber \\
\left(a^{\text{(out)}}_{\vec{k}}\right)^\dagger &= Ta^\dagger_{-\vec{k}}T,
\end{align}
which satisfy harmonic oscillator commutation relations. Inserting the scattering eigenmodes in Eq.~\eqref{eq:eigenmodes} into Eq.~\eqref{eq:time_reversed_electric_field} yields the time-reversed field modes
\begin{align} \label{eq:time_reversed_modes}
\vec{E}^*_{-\vec{k},\lambda}(\vec{r}) = C^*_{-\vec{k},\lambda} \hat{\vec{e}}^*_{-\vec{k},\lambda} \ee^{\ii \vec{k} \vec{r}} + \vec{E}_{-\vec{k},\lambda}^{(s)*}(\vec{r}).
\end{align}
We now see that the time-reversed scattering modes \eqref{eq:time_reversed_modes} are indeed of the form \eqref{eq:output_modes}, since the complex conjugate of an incoming multipole field is an outgoing multipole field. The relevant output modes are therefore the ones described by the operators $a^{\text{(out)}}_{\vec{k}}$, which obey the Heisenberg equations
\begin{align} \label{eq:heisenberg_output}
\dot{a}^{\text{(out)}}_{\vec{k}} =& -\ii\omega a^{\text{(out)}}_{\vec{k}} \nonumber\\
& -\ii\frac{\vec{E}^*_{0}\cdot\vec{d}}{\hbar}\sqrt{\frac{\Gamma}{2\pi}}\frac{g^*(-\hat{\vec{k}})}{\omega-\omega_0+\ii\Gamma/2}\sigma_-.
\end{align}
The important thing to notice now is that the dynamics of $\sigma_-$ can be calculated without refering to any output modes, so that $\sigma_-$ can be treated as an external parameter in Eq.~\eqref{eq:heisenberg_output}. This enables us to make output calculations without having to keep track of an infinite number of operators. This is schematically depicted in Fig.~\ref{fig:output}.

To summarize this section, we have established a way to treat the internal dynamics of an open cavity under arbitrary illumination with only a small number of operators. We have furthermore established a way to relate the results of the internal calculations to the temporal dynamics of individual Fourier components of the far field. This enables us to completely describe optical systems with radiative losses in terms of experimentally accessible quantities. Examples of this procedure are demonstrated in the following Section.

\section{Relation to Jaynes-Cummings Model}
\label{sec:dynamics}

In order to offer a verification of the theory developed above, we now turn to the task of retrieving the well-established Jaynes-Cummings model from our formalism. Consider an input state of the form
\begin{align}
\ket{\Phi} &:= \int d^3k \frac{c_0^{3/2}}{\omega\sqrt{G'}}g'(\hat{\vec{k}})\sqrt{\frac{\Gamma}{2\pi}}\frac{1}{\omega-\omega_0-\ii\Gamma'/2}a^\dagger_{\vec{k}}\ket{0},
\end{align}
where the normalization factor $G'$ is defined as
\begin{align}
G' &:= \int d\Omega_k \left|g'(\hat{\vec{k}})\right|^2.
\end{align}
It is easy to check that the initial state $\ket{\Phi}$ is properly normalized, i.e. $\braket{\Phi|\Phi} = 1$. We will now proceed to derive, from Eqs.~\eqref{eq:complete_heisenberg_a} and \eqref{eq:complete_heisenberg_s}, the Heisenberg equations for the number operators $a^\dagger a$ and $\sigma_+\sigma_-$, as well as the appropriate initial conditions arising from the initial state $\ket{\Phi}$.

We first want to determine the effect of the zero-time noise operators $F_n(0)$ on the initial state $\ket{\Phi}$:
\begin{align} \label{eq:fn_on_phi}
F_n(0)\ket{\Phi} = \int d^3k& \frac{c_0^3}{\omega^2\sqrt{GG'}}g'(\hat{\vec{k}})g^*(\hat{\vec{k}})\frac{\sqrt{\Gamma\Gamma'}}{2\pi} \nonumber \\
& (\omega-\omega_0)^n \frac{1}{\omega-\omega_0-\ii\Gamma'/2} \ket{0}.
\end{align}
Comparing Eq.~\eqref{eq:fn_on_phi} with Eq.~\eqref{eq:fn_deriv_1}, we see that the frequency integrals are formally identical. Since the frequency integral in Eq.~\eqref{eq:fn_deriv_1} vanishes, as is demonstrated in Appendix \ref{app:fn_deriv}, we conclude that
\begin{align}
F_n(0)\ket{\Phi} &= 0.
\end{align}
But since the zero order noise operator $F_0(t)$ is of the form \eqref{eq:solution_f} at all times, we conclude that
\begin{align} \label{eq:f0_on_phi}
F_0(t)\ket{\Phi} &= 0.
\end{align}
Next we consider the action of the cavity operator at zero time on the initial state:
\begin{align} \label{eq:a_on_phi}
a(0)\ket{\Phi} = & \int d\Omega_k \frac{g'(\hat{\vec{k}})}{\sqrt{G'}} \frac{g^*(\hat{\vec{k}})}{\sqrt{G}} \frac{\sqrt{\Gamma\Gamma'}}{2\pi} \nonumber \\
& \int d\omega \frac{1}{\omega-\omega_0-\ii\Gamma'/2} \frac{1}{\omega-\omega_0+\ii\Gamma/2} \ket{0} \nonumber \\
= & \left(\frac{g}{\sqrt{G}} * \frac{g'}{\sqrt{G'}}\right) \frac{\sqrt{\Gamma\Gamma'}}{(\Gamma+\Gamma')/2} \ket{0},
\end{align}
where the scalar product between two angular functions is defined as
\begin{align}
a * b := \int d\Omega_k \left[a(\hat{\vec{k}})\right]^* b(\hat{\vec{k}}).
\end{align}
From Eq.~\eqref{eq:a_on_phi} it is now easy to obtain the initial photon number in the cavity
\begin{align} \label{eq:aa_init}
\bra{\Phi}a^\dagger(0)a(0)\ket{\Phi} &= \left[\frac{g}{\sqrt{G}} * \frac{g'}{\sqrt{G'}}\right]^2 \left[\frac{\sqrt{\Gamma\Gamma'}}{(\Gamma+\Gamma')/2}\right]^2.
\end{align}
Please note that due to the Cauchy-Schwarz inequality \cite{Bronshtein2013}
\begin{align}
\left[\frac{g}{\sqrt{G}} * \frac{g'}{\sqrt{G'}}\right]^2 &\leq \left(\frac{g}{\sqrt{G}} * \frac{g}{\sqrt{G}}\right) \left(\frac{g'}{\sqrt{G'}} * \frac{g'}{\sqrt{G'}}\right) = 1,
\end{align}
and that the geometric mean of two numbers is always smaller than the algebraic mean \cite{Bronshtein2013}
\begin{align}
\frac{\sqrt{\Gamma\Gamma'}}{(\Gamma+\Gamma')/2} &\leq 1.
\end{align}
It therefore follows that the initial number of cavity photons in Eq.~\eqref{eq:aa_init} is smaller than $1$, as is required from the fact that only a single photon is incident.

We now turn to the problem of deriving equations of motion for the number operator expectation values $a^\dagger a$ and $\sigma_+ \sigma_-$. To this end we first use Eqs.~\eqref{eq:complete_heisenberg_a}, \eqref{eq:complete_heisenberg_s} and \eqref{eq:f0_on_phi} to derive
\begin{align} \label{eq:single_photon_dynamics_1}
\frac{\text{d}}{\text{d}t} \braket{a^\dagger a} =& -\Gamma\braket{a^\dagger a} + 2\text{Im}\left[\kappa\braket{a^\dagger \sigma_-}\right], \\
\label{eq:single_photon_dynamics_2}
\frac{\text{d}}{\text{d}t} \braket{\sigma_+ \sigma_-} =& 2\text{Im}\left[\kappa\left(1+\ii\frac{\Gamma}{2\omega_0}\right)\braket{a^\dagger\sigma_z\sigma_-}\right], \\
\label{eq:single_photon_dynamics_3}
\frac{\text{d}}{\text{d}t} \braket{a^\dagger \sigma_-} =& \left[-\ii(\omega_a-\omega_0) - \frac{\Gamma}{2}\right] \braket{a^\dagger\sigma_-} + \nonumber \\
&\ii\kappa^*\braket{\sigma_+ \sigma_-} + \ii\kappa^*\left(1-\ii\frac{\Gamma}{2\omega_0}\right) \braket{\sigma_za^\dagger a},
\end{align}
which does not yet form a closed system of equations. In order to change this, we first use the Pauli operator identity
\begin{align}
\sigma_z\sigma_- &= -\sigma_-,
\end{align}
in order to change $\braket{a^\dagger\sigma_-\sigma_z}$ in Eq.~\eqref{eq:single_photon_dynamics_2} to $-\braket{a^\dagger\sigma_-}$. Next we rewrite the remaining problematic operator as
\begin{align} \label{eq:rewrite_szaa}
\braket{\sigma_za^\dagger a} = 2\braket{a^\dagger\sigma_+\sigma_-a} - \braket{a^\dagger a},
\end{align}
and use the fact that the total excitation number of the original Hamiltonian \eqref{eq:exact_hamiltonian}
\begin{align}
\int d^3k a_{\vec{k}}^\dagger a_{\vec{k}} + \sigma_+ \sigma_-,
\end{align}
is conserved. This means that, for the given initial state, the solution to the Schr\"{o}dinger equation is of the form
\begin{align} \label{eq:schroedinger_formal_solution}
\ket{\Phi(t)} = \int d^3k A(\vec{k}, t)\ket{1_{\vec{k}}, g} + B(t)\ket{0, e},
\end{align}
where $g$ and $e$ refer to the ground and excited state of the atom, respectively. Combining Eqs.~\eqref{eq:rewrite_szaa} and \eqref{eq:schroedinger_formal_solution}, we notice that
\begin{align}
\braket{\sigma_z a^\dagger a} &= -\braket{a^\dagger a}.
\end{align}
Inserting this result into Eq.~\eqref{eq:single_photon_dynamics_3}, we see that the equations of motion are now indeed closed.

\begin{figure*}
\begin{center}
\includegraphics[width=\textwidth]{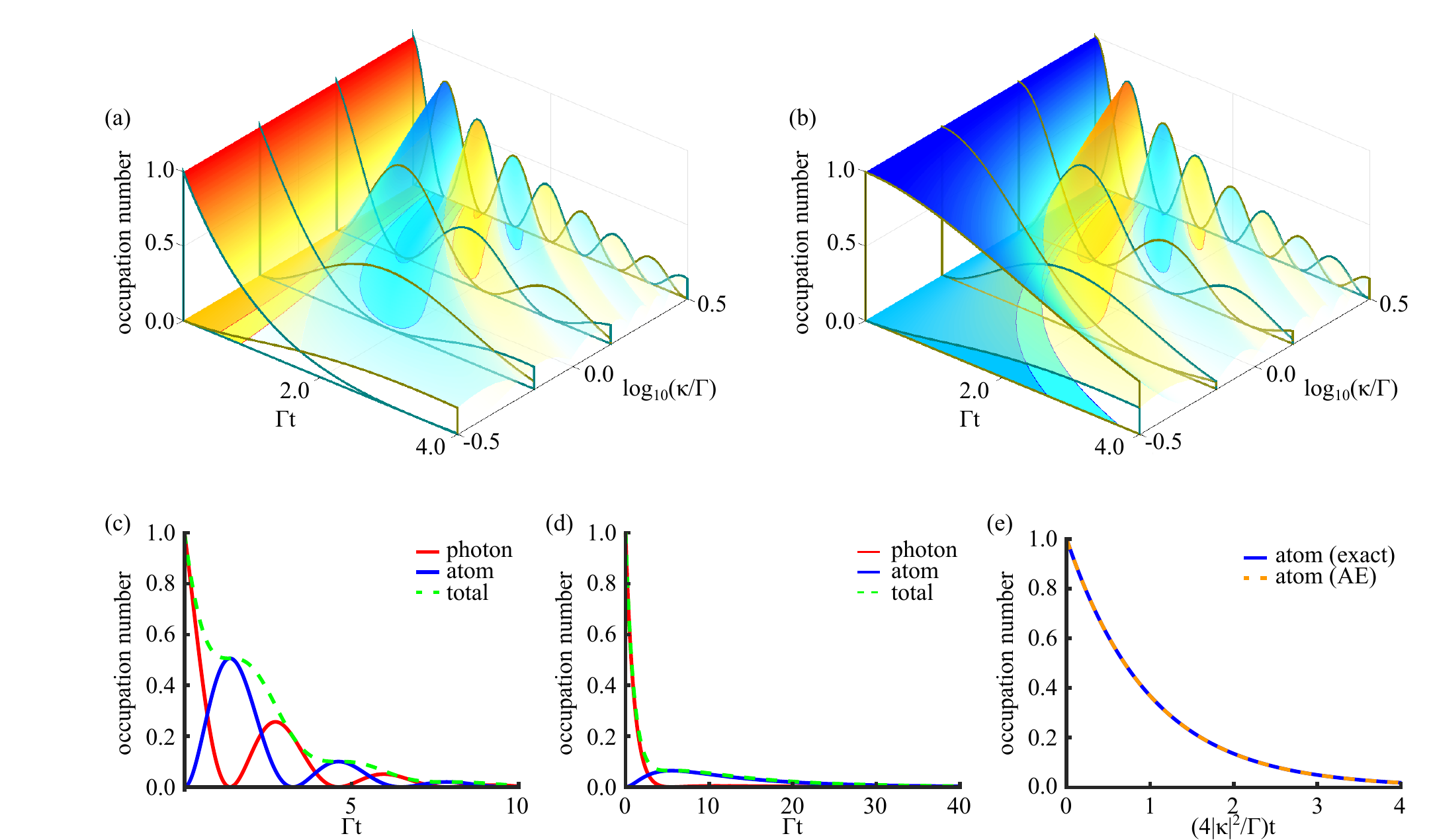}
\caption{Results of numerical simulations of Eqs.~\eqref{eq:single_photon_dynamics_2} with a single-excitation input state: (a) Temporal system dynamics for different values of the coupling-to-loss ratio $\kappa/\Gamma$ when a single photon is injected into the system. Red colors correspond to the photonic and blue colors to the atomic excited state. (b) Same as (a) but with the atom initially excited. (c) Strong coupling case $\kappa = \Gamma = 10^{-3}\omega_0$. The characteristic Rabi oscillations are clearly visible. (d) Weaker coupling $\kappa = 10^{-4}\omega_0$ and $\Gamma = 10^{-3}\omega_0$. Please note the different timescales of photonic and atomic decay. (e) Atomic decay for $\kappa = 10^{-5}\omega_0$ and $\Gamma = 10^{-3}\omega_0$ with the atom initially excited. The solid line refers to the numerical solution of the exact equations of motion and the dashed line is the approximate result obtained from adiabatic elimination (AE).}
\label{fig:3in1}
\end{center}
\end{figure*}

The above equations of motion are very similar to the ones derived from the Jaynes-Cummings (JC) Hamiltonian in the single photon case, except for the damping terms and the small asymmetry in the light-matter coupling terms. We therefore expect to find the well-known phenomena of the JC model when solving them. In order to see if these expectations are accurate, we present a number of numerical simulations for different system parameters. We assume without loss of generality that $g' = g$ and $\Gamma' = \Gamma$, so that the initial photon number in the cavity is exactly $1$. Figure~\ref{fig:3in1}(a) shows the result for a system in the strong coupling regime, i.e. $\kappa > \Gamma/2$. The temporal oscillations of energy between electromagnetic field and atom, known as Rabi oscillations, are clearly visible. Due to the dissipative nature of the cavity mode the oscillations are enveloped by a monotonically decreasing function, which represents the total number of (electromagnetic and atomic) excitations in the system. The step-like shape of the total excitation can be explained by recalling that only the electromagnetic mode is dissipative. Hence, whenever all of the energy is stored in the atom, no dissipation occurs and the total excitation number has a plateau. The result for a weakly coupled system is shown in Fig.~\ref{fig:3in1}(b). The cavity photon is seen to decay rapidly, exciting the atom only weakly. The atomic excitation then also starts decaying, although the atom itself is non-dissipative. This is, of course, due to the atom being coupled to the dissipative cavity.

In order to further investigate the decay of the atom via the cavity, we turn to the case where the atom is initially excited, while the electromagnetic field is in its ground state:
\begin{align} \label{eq:atom_init}
\ket{\Phi} &= \ket{0, e}.
\end{align}
One can easily convince oneself that the equations of motion derived above still hold for the initial state in Eq.~\eqref{eq:atom_init}. The initial conditions, however, have to be changed to $\braket{\sigma_+\sigma_-}=1$ and $\braket{a^\dagger a} = 0$. If we restrict ourselves to the case $\kappa \ll \Gamma$, we can obtain an approximate solution by adiabatic elimination (AE) of the cavity mode \cite{Fewell2005}. The AE result reads
\begin{align}
\label{eq:ae_atom}
\braket{\sigma_+(t)\sigma_-(t)} &= \exp\left(-4\frac{\left|\kappa\right|^2}{\Gamma}t\right).
\end{align}
The numerical result for the complete set of equations of motion is shown in Fig.~\ref{fig:3in1}(c) and compared to the AE result. We see an excellent agreement between the approximate and the complete solution. The deexcitation of the atom to due to its coupling to the electromagnetic vacuum is, of course, just the well-known phenomenon of spontaneous emission. The dependence of the emission time on the coupling strength is a manifestation of the Purcell effect \cite{Purcell1946}.

If the decay of an initially excited atom is indeed due to spontaneous emission, then the energy should be transferred to the output modes described in Sec.~\ref{sec:io}. In order to verify this, we first derive from Eqs.~\eqref{eq:complete_heisenberg_a}, \eqref{eq:complete_heisenberg_s} and \eqref{eq:heisenberg_output} the single-photon equations of motion
\begin{align}
\label{eq:heisenberg_output2}
\frac{d}{dt} \braket{a^{\text{(out)}\dagger}_{\vec{k}} a^{\text{(out)}}_{\vec{k}}} =& 2 \text{Im}\left[\zeta_{\vec{k}} \braket{a^{\text{(out)}\dagger}\sigma_-}\right], \\
\label{eq:heisenberg_output_s}
\frac{d}{dt} \braket{a^{\text{(out)}\dagger}_{\vec{k}}\sigma_-} =& -\ii \Delta_{\vec{k}} \braket{a^{\text{(out)}\dagger}_{\vec{k}}\sigma_-} \nonumber \\
& + \ii \zeta^*_{\vec{k}} \braket{\sigma_+\sigma_-} -\ii \kappa^* \braket{a^{\text{(out)}\dagger}_{\vec{k}}a}, \\
\label{eq:heisenberg_output_a}
\frac{d}{dt} \braket{a^{\text{(out)}\dagger}_{\vec{k}}a} =& \left(-\ii\Delta_{\vec{k}} - \frac{\Gamma}{2}\right) \braket{a^{\text{(out)}\dagger}_{\vec{k}}a} \nonumber \\
& + \ii \zeta^*_{\vec{k}} \braket{\sigma_+a} - \ii\kappa \braket{a^{\text{(out)}\dagger}_{\vec{k}}\sigma_-},
\end{align}
where the definitions
\begin{align}
\label{eq:zeta}
\zeta_{\vec{k}} :=& \frac{\vec{E}_0^* \cdot \vec{d}}{\hbar} \sqrt{\frac{\Gamma}{2\pi}} \frac{g^*(-\vec{k})}{\omega_{\vec{k}}-\omega_0+\ii\Gamma/2}, \\
\Delta_{\vec{k}} :=& \omega_0 - \omega_{\vec{k}},
\end{align}
were used. Since we assume $\kappa \ll \Gamma$, Eq.~\eqref{eq:heisenberg_output_a} can be adiabatically eliminated. Equations~\eqref{eq:heisenberg_output_s} then become
\begin{align}
\label{eq:ae_output_1}
\frac{d}{dt} \braket{a^{\text{(out)}\dagger}_{\vec{k}}\sigma_-} =& \left[-\ii\Delta_{\vec{k}} - \frac{\left|\kappa\right|^2}{\ii\Delta_{\vec{k}}+\Gamma/2}\right] \braket{a^{\text{(out)}\dagger}_{\vec{k}}\sigma_-} \nonumber \\
& +\ii \zeta^*_{\vec{k}} \braket{\sigma_+\sigma_-} + \zeta^*_{\vec{k}} \frac{\kappa^*}{\ii\Delta_{\vec{k}}+\Gamma/2}\braket{\sigma_+a}.
\end{align}
This equation can be further simplified, if one uses the AE result
\begin{align}
\braket{\sigma_+a} \approx -\ii \frac{\kappa}{\Gamma/2} \braket{\sigma_+\sigma_-} \ll \braket{\sigma_+\sigma_-}.
\end{align}
Equation \eqref{eq:ae_output_1} can then be solved to yield
\begin{align}
\label{eq:ae_output_2}
&\braket{a^{\text{(out)}\dagger}_{\vec{k}}\sigma_-}_t = \ii \zeta^*_\vec{k} \int_0^t dt' \nonumber\\
&\exp\left(\left[-\ii\Delta_{\vec{k}}-\frac{\left|\kappa\right|^2}{\ii\Delta_{\vec{k}}+\Gamma/2}\right]\left[t-t'\right]\right) \braket{\sigma_+\sigma_-}_{t'},
\end{align}
where the indices of the expectation values denote the time of evaluation. Inserting Eq.~\eqref{eq:ae_output_2} into Eq.~\eqref{eq:heisenberg_output2} and integrating the resulting equation leads to
\begin{align}
\label{eq:ae_output3}
&\braket{a^{\text{(out)}\dagger}_{\vec{k}}a^{\text{(out)}}_{\vec{k}}} = 2\left|\zeta_{\vec{k}}\right|^2 \text{Re}\left\{\int_0^tdt' \int_0^{t'}dt'' \right.\nonumber \\
&\left.\exp\left[\left(-\ii\Delta_{\vec{k}}-\frac{\left|\kappa\right|^2}{\ii\Delta_{\vec{k}}+\Gamma/2}\right)\left(t' - t''\right)\right] \braket{\sigma_+\sigma_-}_{t''}\right\}.
\end{align}
Using the AE result Eq.~\eqref{eq:ae_atom}, the integrations in Eq.~\eqref{eq:ae_output3} can be easily performed. The result reads
\begin{widetext}
\begin{align}
\label{eq:output2_original}
&\braket{a^{\text{(out)}\dagger}_{\vec{k}}a^{\text{(out)}}_{\vec{k}}} = \left|\zeta_{\vec{k}}\right|^2\text{Re}\left\{ \frac{\Gamma/2}{\left|\kappa\right|^2} \frac{1}{\ii\Delta_{\vec{k}}+\left|\kappa\right|^2/(\ii\Delta_{\vec{k}}+\Gamma/2)-2\left|\kappa\right|^2/(\Gamma/2)} \left[1-\exp\left(-2\frac{\left|\kappa\right|^2}{\Gamma/2}t\right)\right] \right. \nonumber \\
&\left.+ 2 \frac{1}{-\ii\Delta_{\vec{k}}-\left|\kappa\right|^2/(\ii\Delta_{\vec{k}}+\Gamma/2)}\frac{1}{\ii\Delta_{\vec{k}}+\left|\kappa\right|^2/(\ii\Delta_{\vec{k}}+\Gamma/2)-2\left|\kappa\right|^2/(\Gamma/2)} \left[1 - \exp\left(-\ii\Delta_{\vec{k}}-\frac{\left|\kappa\right|^2}{\ii\Delta_{\vec{k}}+\Gamma/2}t\right)\right] \right\}.
\end{align}
\end{widetext}
Considering the denominators in Eq.~\eqref{eq:output2_original}, it becomes clear that the number of photons with frequency detuning $\Delta_{\vec{k}}$ is small, unless $\Delta_{\vec{k}} \lesssim \left|\kappa\right|^2/(\Gamma/2)$. But due to the adiabatic assumption this means $\Delta_{\vec{k}} \ll \Gamma/2$, which allows for the following simplification of Eq.~\eqref{eq:output2_original}:
\begin{align}
\label{eq:output_result}
&\braket{a^{\text{(out)}\dagger}_{\vec{k}}a^{\text{(out)}}_{\vec{k}}} \approx \left|\zeta_{\vec{k}}\right|^2 \frac{1}{\Delta_{\vec{k}}^2 + \left[\left|\kappa\right|^2/\left(\Gamma/2\right)\right]^2} \nonumber \\
&\left[1-2\cos\left(\Delta_{\vec{k}}t\right)\exp\left(-\frac{\left|\kappa\right|^2}{\Gamma/2}t\right)+\exp\left(-2\frac{\left|\kappa\right|^2}{\Gamma/2}t\right)\right].
\end{align}
Likewise, the expression \eqref{eq:zeta} for $\zeta_{\vec{k}}$ can be simplified to yield
\begin{align}
\label{eq:zeta_simplified}
\left|\zeta_{\vec{k}}\right|^2 \approx& \left|\frac{\vec{E}_0 \cdot \vec{d}}{\hbar}\right| \frac{\Gamma}{2\pi} \frac{\left|g(-\hat{\vec{k}})\right|^2}{(\Gamma/2)^2} \nonumber \\
=& \frac{c_0^3}{\omega_0^2}\frac{\left|g(-\hat{\vec{k}})\right|^2}{G} \frac{\left|\kappa\right|^2}{\pi(\Gamma/2)}.
\end{align}
To acquire the total output photon number, the result in Eq.~\eqref{eq:output_result} has to be integrated over the wave vector. Using Eq.~\eqref{eq:zeta_simplified} and writing the cosine in terms of exponential functions, one finds
\begin{widetext}
\begin{align}
\int d^3k\braket{a^{\text{(out)}\dagger}_{\vec{k}}a^{\text{(out)}}_{\vec{k}}}
 \approx& \int_{-\infty}^\infty d\omega \frac{\omega}{\omega_0} \frac{\left|\kappa\right|^2}{\pi\Gamma/2} \frac{1}{\Delta_{\vec{k}}^2+\left[\left|\kappa\right|^2/\left(\Gamma/2\right)\right]} \nonumber \\
& \left[1+\exp\left(-2\frac{\left|\kappa\right|^2}{\Gamma/2}t\right)-\exp\left(-\frac{\left|\kappa\right|^2}{\Gamma/2}t+\ii\Delta_{\vec{k}}t\right)-\exp\left(-\frac{\left|\kappa\right|^2}{\Gamma/2}t-\ii\Delta_{\vec{k}}t\right)\right] \nonumber \\
=&\left[1 - \exp\left(-2\frac{\left|\kappa\right|^2}{\Gamma/2}t\right)\right] = 1-\braket{\sigma_+\sigma_-}_t,
\end{align}
\end{widetext}
where the integrations are performed with standard contour integral techniques. We therefore find that the total excitation number is a constant of motion and that the number of photons asymptotically reaches $1$ for large times. This is in perfect agreement with the requirements of the original Hamiltonian \eqref{eq:exact_hamiltonian}, as well as the physical intuition regarding spontaneous emission.

\section{Conclusion}
\label{sec:conclusion}

We demonstrated the extenstion of the unitary internal cavity dynamics in QED by a rigorous quantum description of radiative losses. Unlike the canonical formulation based on a phenomenological coupling to a thermal bath, we have derived a description employing a chain of noise operators. Furthermore, we added input and output channels to the formalism that allow for a complete description of the dynamics: starting from incident far field illumination, incorporating all unitary cavity related processes, and culminating in far field emission. The procedure suggested here consists of the following steps:
\begin{itemize}
\item Characterize the cavity classically by determining the resonance frequency and linewidth as well as the dependence on the illumination direction.
\item Calculate the light-matter coupling constant from the properly normalized field strength (according to \cite{Oppermann2017}), the cavity parameters and the emitter's transition dipole moment.
\item Evaluate $F_0(t)$ by means of the zero-time noise operators $F_n(0)$ and the initial photonic state.
\item Calculate the internal dynamics of the cavity mode.
\item Solve the equations of motion for the output modes of interest.
\end{itemize}
We discussed single-photon dynamics in a leaky cavity coupled to a single atom and retrieved the familiar Jaynes-Cummings model, but with the added possibility of calculating the far field dynamics. However, the formalism presented here can be employed to describe a multitude of different scenarios of light-matter interaction, which go beyond the simple Jaynes-Cummings model. We hope that this work will pave the way towards a more rigorous description of open optical cavities and their interaction with the far field.

\section*{Acknowledgments}
\label{sec:acknowledgments}
The study was supported by the Karlsruhe School of Optics and Photonics (KSOP). The authors also wish to thank the Deutscher Akademischer Austauschdienst (PPP Poland) and the Ministry of Science and Higher Education in Poland.

\appendix

\section{Derivation of System Operator Heisenberg Equations}
\label{app:as_deriv}

Using the definition \eqref{eq:cavity_mode_definition} together with the Heisenberg equation of motion \eqref{eq:exact_heisenberg_a}, one derives:
\begin{align}
\dot{a} =& \int d^3k \frac{c_0^{3/2}}{\sqrt{G}\omega} \sqrt{\frac{\Gamma}{2\pi}} \frac{g^*(\hat{\vec{k}})}{\omega-\omega_0+\ii\Gamma/2} \dot{a}_{\vec{k}} \nonumber \\
=& \int d^3k \frac{c_0^{3/2}}{\sqrt{G}\omega} \sqrt{\frac{\Gamma}{2\pi}} \frac{g^*(\hat{\vec{k}})}{\omega-\omega_0+\ii\Gamma/2} (-\ii\omega) a_{\vec{k}} \nonumber \\
&-\ii\frac{\vec{E}_0 \cdot \vec{d}}{\hbar} \sigma_-\int d^3k \frac{c_0^{3/2}}{\sqrt{G}\omega} \frac{\Gamma}{2\pi} \frac{\left|g(\hat{\vec{k}})\right|^2}{\left(\omega-\omega_0\right)^2+\left(\Gamma/2\right)^2} \nonumber \\
=& -\ii\int d^3k \frac{c_0^{3/2}}{\sqrt{G}\omega} \sqrt{\frac{\Gamma}{2\pi}} g^*(\hat{\vec{k}}) a_{\vec{k}} \nonumber \\
& \phantom{-\ii\int d^3k \frac{c_0^{3/2}}{\sqrt{G}\omega}} \frac{\omega-\omega_0+\ii\Gamma/2+\omega_0-\ii\Gamma/2}{\omega-\omega_0+\ii\Gamma/2} \nonumber \\
&-\ii\frac{\sqrt{G}}{c_0^{3/2}}\frac{\vec{E}_0 \cdot \vec{d}}{\hbar} \sigma_-\int_0^\infty d\omega  \frac{\Gamma}{2\pi} \frac{\omega-\omega_0+\omega_0}{\left(\omega-\omega_0\right)^2+\left(\Gamma/2\right)^2}.
\end{align}
Since we assume $\Gamma \ll \omega_0$, the lower integration boundary in the second term can be approximately shifted to $-\infty$. Noticing that the part of the second integral antisymmetric in $\omega-\omega_0$  vanishes and splitting up the fracture under the first integral, one arrives at
\begin{align}
\dot{a} = & -\ii\int d^3k \frac{c_0^{3/2}}{\sqrt{G}\omega} \sqrt{\frac{\Gamma}{2\pi}} g^*(\hat{\vec{k}}) a_{\vec{k}} \nonumber \\
&+ (-\ii\omega_0 - \frac{\Gamma}{2}) \int d^3k \frac{c_0^{3/2}}{\sqrt{G}\omega} \sqrt{\frac{\Gamma}{2\pi}} \frac{g^*(\hat{\vec{k}})}{\omega-\omega_0+\ii\Gamma/2} a_{\vec{k}} \nonumber \\
&-\ii\frac{\sqrt{G}}{c_0^{3/2}}\frac{\vec{E}_0 \cdot \vec{d}}{\hbar} \sigma_-.
\end{align}
Employing the definitions in Eqs.~\eqref{eq:cavity_mode_definition}, \eqref{eq:coupling_constant} and \eqref{eq:fluctuation_definition} this becomes
\begin{align}
\dot{a} &\approx (-i\omega_0-\Gamma/2)a - \ii\kappa\sigma_- -\ii F_0.
\end{align}

Turning now to the atom dynamics, the equation of motion \eqref{eq:exact_heisenberg_s} can be written
\begin{align}
\dot{\sigma}_- =& -\ii\omega_{a}\sigma_- \nonumber\\
& +\ii\frac{\vec{E}^*_{0}\cdot\vec{d}^*}{\hbar}\int \frac{d^3k}{\omega} \sqrt{\frac{\Gamma}{2\pi}}\frac{g^*(\hat{\vec{k}})\omega}{\omega-\omega_0+\ii\Gamma/2}\sigma_za_{\vec{k}} \nonumber\\
=& -\ii\omega_{a}\sigma_- \nonumber\\
& +\ii\frac{\vec{E}^*_{0}\cdot\vec{d}^*}{\hbar}\int \frac{d^3k}{\omega} \sqrt{\frac{\Gamma}{2\pi}}g^*(\hat{\vec{k}}) \sigma_za_{\vec{k}} \nonumber \\
&\phantom{+\ii\frac{\vec{E}^*_{0}\cdot\vec{d}^*}{\hbar}\int \frac{d^3k}{\omega}}\frac{\omega-\omega_0+\ii\Gamma/2+\omega_0-\ii\Gamma/2}{\omega-\omega_0+\ii\Gamma/2} \nonumber\\
=& -\ii\omega_{a}\sigma_- \nonumber\\
& +\ii\frac{\vec{E}^*_{0}\cdot\vec{d}^*}{\hbar}\sigma_z\int \frac{d^3k}{\omega} \sqrt{\frac{\Gamma}{2\pi}}g^*(\hat{\vec{k}})a_{\vec{k}} \nonumber\\
& + \ii \left(\omega_0-\ii\frac{\Gamma}{2}\right) \frac{\vec{E}^*_{0}\cdot\vec{d}^*}{\hbar}\sigma_z \nonumber \\
&\phantom{+}\int \frac{d^3k}{\omega} \sqrt{\frac{\Gamma}{2\pi}}\frac{g(\hat{\vec{k}})}{\omega-\omega_0+\ii\Gamma/2}a_{\vec{k}}.
\end{align}
Using once again the definitions in Eqs.~\eqref{eq:cavity_mode_definition}, \eqref{eq:coupling_constant} and \eqref{eq:fluctuation_definition} this can be written
\begin{align}
\dot{\sigma}_- =& -\ii\omega_{a}\sigma_- \nonumber\\
& +\ii\frac{\vec{E}^*_{0}\cdot\vec{d}^*}{\hbar}\int d^3k \sqrt{\frac{\Gamma}{2\pi}}\frac{g^*(\hat{\vec{k}})}{\omega-\omega_0+\ii\Gamma/2}\sigma_za_{\vec{k}}.
\end{align}

\section{Derivation of Noise Operator Heisenberg Equations}
\label{app:fn_deriv}

Using the definition in Eq.~\eqref{eq:fluctuation_definition} together with the Heisenberg equations \eqref{eq:exact_heisenberg_a} and \eqref{eq:exact_heisenberg_s} we arrive at
\begin{align} \label{eq:fn_deriv_1}
\dot{F}_n =& -\ii\omega_0 F_n -\ii F_{n+1} \nonumber \\
&-\ii \frac{\kappa}{\omega_0} \frac{\Gamma}{2\pi} \sigma_- \int_0^\infty d\omega \omega(\omega-\omega_0)^n \frac{1}{\omega-\omega_0-\ii\Gamma/2}.
\end{align}
As can be easily seen, the above frequency integral is highly divergent. This is due to the fact that we assumed a perfect Lorentzian frequency dependence of the electromagnetic field at the emitter position. In a real system, however, one would not expect this assumption to hold for frequencies far off-resonance. Especially for very high frequencies one expects rapid oscillations of the field strength, so that the high frequency contributions average out to zero. Equation \eqref{eq:lorentzian_mode} therefore has to be modified to take the off-resonance contributions into account. We do this by adding an Gaussian envelope that decays on time scales large compared to the Lorentzian linewidth $\Gamma$, but small compared to $\omega_0$:
\begin{align} \label{eq:lorentzian_mode_modified}
&\vec{E}_{\vec{k},\lambda}(\vec{r}_{a}) = \vec{E}_{0} \delta_{\lambda, \lambda_0} \sqrt{\frac{\Gamma}{2\pi}}\frac{g(\hat{\vec{k}})\ee^{-(\omega-\omega_0)^2/\beta^2}}{\omega-\omega_0-\ii\Gamma/2}, \\
&\Gamma \ll \beta \ll \omega_0. \nonumber
\end{align}
Using Eq.~\eqref{eq:lorentzian_mode_modified} instead of Eq.~\eqref{eq:lorentzian_mode}, the integral in Eq.~\eqref{eq:fn_deriv_1} becomes
\begin{align} \label{eq:integral_infinite}
&\int_0^\infty d\omega \omega(\omega-\omega_0)^n \frac{\ee^{-(\omega-\omega_0)^2/\beta^2}}{\omega-\omega_0+\ii\Gamma/2} \nonumber\\
&\approx\int_{-\infty}^\infty d\omega \omega(\omega-\omega_0)^n \frac{\ee^{-(\omega-\omega_0)^2/\beta^2}}{\omega-\omega_0+\ii\Gamma/2},
\end{align}
where the lower integration boundary has been approximately extended to $-\infty$, since the exponential function decays much faster than any polynomial can grow.

The integral in Eq.~\eqref{eq:integral_infinite} can now be solved by contour integration techniques, if one introduces an auxiliary factor of $\exp(\pm\ii\epsilon\omega)$. But while $\epsilon$ can just be chosen to be infinitesimally small, the choice of sign in the exponent leads to very different results. This is due to the fact that the integrand only possesses a pole in the upper half-plane. Hence, in order to find a meaningful result we need to eliminate one of the two possibilities by physical reasoning. This is similar to choosing the retarded instead of the advanced Green's function, since the later violates causality. However, in the current case it is not immediately obvious which solution is the unphysical one. Only once we obtain the solutions for both possible equations will it be obvious which one to choose. For this reason we consider, for the moment, both possible solutions:
\begin{align} \label{eq:physical_fn}
\dot{F}_n =& -\ii\omega_0 F_n -\ii F_{n+1},~\text{(lower half-plane)}\\
\label{eq:unphysical_fn}
\dot{F}_n =& -\ii\omega_0 F_n -\ii F_{n+1}\nonumber\\
&-\ii\kappa\left[1+\ii\frac{\Gamma}{2\omega_0}\right]2\left(\ii\frac{\Gamma}{2}\right)^{n+1}.~\text{(upper half-plane)}
\end{align}
The details of solving both of these equations will be presented in Appendix \ref{app:fn_sol}, where we show that the solution for integration over the lower half-plane is the physical one.

\section{Solution of Noise Operator Heisenberg Equations}
\label{app:fn_sol}

We start by considering the equation of motion \eqref{eq:unphysical_fn} in order to demonstrate its unphysical nature. Formally solving and then iterating the equation leads to
\begin{align} \label{eq:unphysical_iter}
F_n(t) =& \ee^{-\ii\omega_0 t} \sum_{n=0}^\infty \frac{(-\ii t)^n}{n!}F_n(0) \nonumber \\
& + \Gamma \kappa\left[1+\ii\frac{\Gamma}{2\omega_0}\right] \sum_{n=0}^\infty (\frac{\Gamma}{2})^{n} J_n(t;t),
\end{align}
where the operator valued terms $J_n(t)$ read
\begin{align}
J_n(t;t_0) &= \int_0^t dt_1 \ldots \int_0^{t_n}dt_{n+1} \ee^{\ii\omega_0(t_{n+1}-t_0)}\sigma_-(t_{n+1}).
\end{align}
We can now use induction to calculate the values of $J_n(t)$. Since $J_0(t)$ is of the form
\begin{align}
J_0(t;t) = \int_0^t dt_1\ee^{\ii\omega_0(t_{1}-t)}\sigma_-(t_{1}),
\end{align}
the following induction hypothesis is consistent with the base case:
\begin{align}
J_n(t;t_0) = \int_0^t dt' \ee^{\ii\omega_0(t'-t_0)}\sigma_-(t') \frac{(t-t')^{n}}{n!}.
\end{align}
Performing the induction step is now straightforward
\begin{align} \label{eq:j_sol}
J_{n+1}(t;t) =& \int_0^t dt'' J_n(t'';t) \nonumber \\
=& \int_0^t dt''\int_0^{t''} dt' \ee^{\ii\omega_0(t'-t)}\sigma_-(t') \frac{(t''-t')^{n}}{n!} \nonumber \\
=&\int_0^t dt' \ee^{\ii\omega_0(t'-t)} \sigma_-(t') \nonumber \\
& \int_0^{t} dt'' \frac{(t''-t')^{n}}{n!} \Theta(t''-t') \nonumber \\
=&\int_0^t dt' \ee^{\ii\omega_0(t'-t)} \sigma_-(t') \int_{t'}^{t} dt'' \frac{(t''-t')^{n}}{n!} \nonumber \\
=&\int_0^t dt' \ee^{\ii\omega_0(t'-t)} \sigma_-(t') \frac{(t-t')^{n+1}}{(n+1)!},
\end{align}
which is of the required form. Inserting Eq.~\eqref{eq:j_sol} into Eq.~\eqref{eq:unphysical_iter} now yields
\begin{align} \label{eq:unphysical_sol}
F_n(t) =& \ee^{-\ii\omega_0 t} \sum_{n=0}^\infty \frac{(-\ii t)^n}{n!}F_n(0) \nonumber \\
& + \kappa\left[1+\ii\frac{\Gamma}{2\omega_0}\right] \Gamma\int_0^t dt' \ee^{(-\ii\omega_0+\Gamma/2)(t-t')} \sigma_-(t') ,
\end{align}
where the infinite sum was performed to yield an exponential function. Close inspection of Eq.~\eqref{eq:unphysical_sol} reveals that the second term is divergent in time due to the factor $\exp[(\Gamma/2)t]$, which can be pulled in front of the integral. But this would mean that the noise operators grow without limit, driving the temperature of the system towards infinity. The equations of motion \eqref{eq:unphysical_fn} is therefore clearly unphysical.

Turning to \eqref{eq:physical_fn} we find the formal solution
\begin{align} \label{eq:formal_sol_fn}
F_n(t) &= \ee^{-\ii\omega_0 t} F_n(0) - \ii \int_0^t dt' \ee^{-\ii\omega_0 (t-t')}F_{n+1}(t').
\end{align}
Iteration of Eq.~\eqref{eq:formal_sol_fn} then leads to the form
\begin{align} \label{eq:iteration_fn}
F_n(t) &= \ee^{-\ii\omega_0 t} \sum_{m=0}^\infty F_{n+m}(t) (-\ii)^m I_m(t), \\
I_m(t) &=  \int_0^t dt_1 \ldots \int_0^{t_{m-1}} dt_m.
\end{align}
The elements of the series $I_m(t)$ can be easily calculated by induction. First we notice that the base case
\begin{align}
I_0(t) &= 1,
\end{align}
is in agreement with the assumption
\begin{align}
I_m(t) &= \frac{t^m}{m!}.
\end{align}
We now proceed with the inductive step
\begin{align} \label{eq:fn_series_sol}
I_{m+1}(t) &= \int_0^t dt_1 I_m(t_1) = \int_0^tdt_1 \frac{t_1^m}{m!} = \frac{t^{m+1}}{(m+1)!},
\end{align}
hence proving our assumption. Substitution of Eq.~\eqref{eq:fn_series_sol} into Eq.~\eqref{eq:iteration_fn} gives the final result
\begin{align} \label{eq:fn_free_part}
F_n(t) &= \ee^{-\ii\omega_0 t} \sum_{m=0}^\infty \frac{(-\ii t)^m}{m!} F_{n+m}(t).
\end{align}

\bibliography{LCQEDrefs}

\end{document}